\documentclass[proof]{wileyASNA}

\articletype{PROCEEDING}%

\received{26 April 2016}
\revised{6 June 2016}
\accepted{6 June 2016}
\usepackage{float}
\usepackage{graphicx}

\begin{document}

\title{Optical and cm follow-ups of the Changing-Look event in Mkn 590}

\author[1]{Biswaraj Palit*}

\author[2]{Abhijeet Borkar}

\author[1]{Agata R{\'o}\.za\'nska}

\author[1]{Alex Markowitz}

\author[2]{Marzena \'Sniegowska}

\author[3,4]{Swayamtrupta Panda}

\author[5]{David Homan}

\author[1]{Krystian I\l{}kiewicz}

\authormark{Palit \textsc{et al}}

\address[1]{\orgname{Nicolaus Copernicus Astronomical Center, Polish Academy of Science}, \orgaddress{\state{ul. Bartycka 18, Warsaw}, \country{Poland}}}

\address[2]{\orgname{Astronomical Institute of the Czech Academy of Sciences}, \orgaddress{\state{Bo\v{c}n\'i II 1401, CZ-14100 Prague}, \country{Czechia}}}

\address[3]{Gemini Science Fellow}
\address[4]{\orgname{International Gemini Observatory/NSF NOIRLab}, \orgaddress{\state{Casilla 603, La Serena}, \country{Chile}}}
\address[5]{\orgname{Institute of Astronomy, University of Cambridge}, \orgaddress{\state{Madingley Road Cambridge CB3 0HA}, \country{United Kingdom}}}
\corres{*Biswaraj Palit, Nicolaus Copernicus Astronomical Center, ul. Bartycka 18, Warsaw, Poland. \email{bpalit@camk.edu.pl}}


\abstract{The Changing-Look active galactic nucleus Mkn 590 is currently in a rejuvenated state, exhibiting a contemporaneous flux rise across X-rays, UV, optical and cm wavelengths. In this study, we present three new optical spectra obtained with the Nordic Optical Telescope, alongside three 1.4~GHz continuum measurements from the Giant Meterwave Radio Telescope, acquired since Nov. 2024. We identified a clear increase in the broad hydrogen Balmer line emission in the most recent observational epochs. Additionally, the core radio flux densities appear to track the overall X-ray variability, suggesting a possible connection between the accretion flow and jet activity. Based on these data, we aim to explore the evolution of the circumnuclear gas in this source and potential links between accretion and ejection activity.}
\keywords{Active Galactic Nuclei, Changing-Look, Accretion flow, Broad Line Region, Jets}



\maketitle


\section{Introduction}\label{sec1}

Seyfert galaxies host persistently accreting supermassive black holes (SMBHs), known as active galactic nuclei (AGNs). These powerful engines influence galaxy evolution via mechanical and radiative feedback and are major contributors to the unresolved cosmic X-ray background.
 In the past decade, a new class of AGN called Changing-Look AGNs (CL-AGNs) has emerged \citep{2023NatAs...7.1282R}. These sources exhibit dramatic luminosity variations on timescales of just 5–10 years \citep{Panda_Sniegowska_2024ApJS..272...13P}, leading to corresponding changes in the broad components of the hydrogen Balmer emission lines far shorter than the $\sim$10$^6$ year duty cycles predicted by standard accretion theory \citep{2015MNRAS.451.2517S}. 
Among CL-AGNs, a sub-category of sources are classified as Changing-State AGNs (CS-AGNs) where the changes in luminosity are likely caused by rapid shifts in mass accretion rate onto the SMBH, likely causing structural changes in the inner accretion flow. These changes can strongly affect the net ionizing radiation and, consequently, the illumination of the Broad Line Region (BLR). The resulting variations in BLR emission can lead to the observed appearance or disappearance of broad Balmer lines in the optical band, producing transitions between Seyfert types i.e., from a dim, type 2 state to a bright, type 1 state. While structural evolution of the BLR is possible, such changes likely occur on longer timescales than those observed in many CL-AGNs.   

In a few well-monitored CS-AGNs such as Mrk 1018 \citep{saha2025A&A...699A.205S}, NGC 1566 \citep{trip2022ApJ...930..117T}, 1ES 1927+654 \citep{ghoshRitest2023ApJ...955....3G}, and Mrk 590 \citep{lawther2025MNRAS.539..501L,2025MNRAS.540L..14P}, the CL events have been associated with significant modulation of the soft X-ray excess component. This excess is typically attributed to warm, optically thick plasma located above the cold disk near the SMBH \citep{rozanska2015,POP2018A&A...611A..59P,palit2024A&A...690A.308P}. Furthermore, intermittent radio activity coincident with episodes of rising X-ray flux has been detected in at least two CL-AGNs: NGC 1566, which exhibited a significant increase in the millimeter continuum emission \citep{janaNGC2025A&A...699A..62J}, and 1ES 1927+654, where the observations using \textit{K}-band imaging from the Very Large Baseline Array (VLBA) revealed the emergence of bipolar jets following a CL event \citep{meyer2025ApJ...979L...2M}. Additionally, tidal disruption events (TDEs), where stars are tidally stripped apart by the SMBH, also frequently show radio flares within 100-1000 days after an episode of sudden accretion activity \citep{horesh2021NatAs...5..491H,Cendes2024ApJ...971..185C,goodwin2024MNRAS.528.7123G}. These sequence of events suggests a possible connection between accretion driven CL phenomena and episodic jets or outflows launched during spectral state transitions, that are otherwise routinely observed in X-ray binaries \citep[XRBs;][]{fender2004MNRAS.355.1105F}. Together, these systems offer a rare window into the coupled accretion-ejection processes in SMBHs.

\begin{figure*}[!]
    \centering
    \includegraphics[scale=.5]{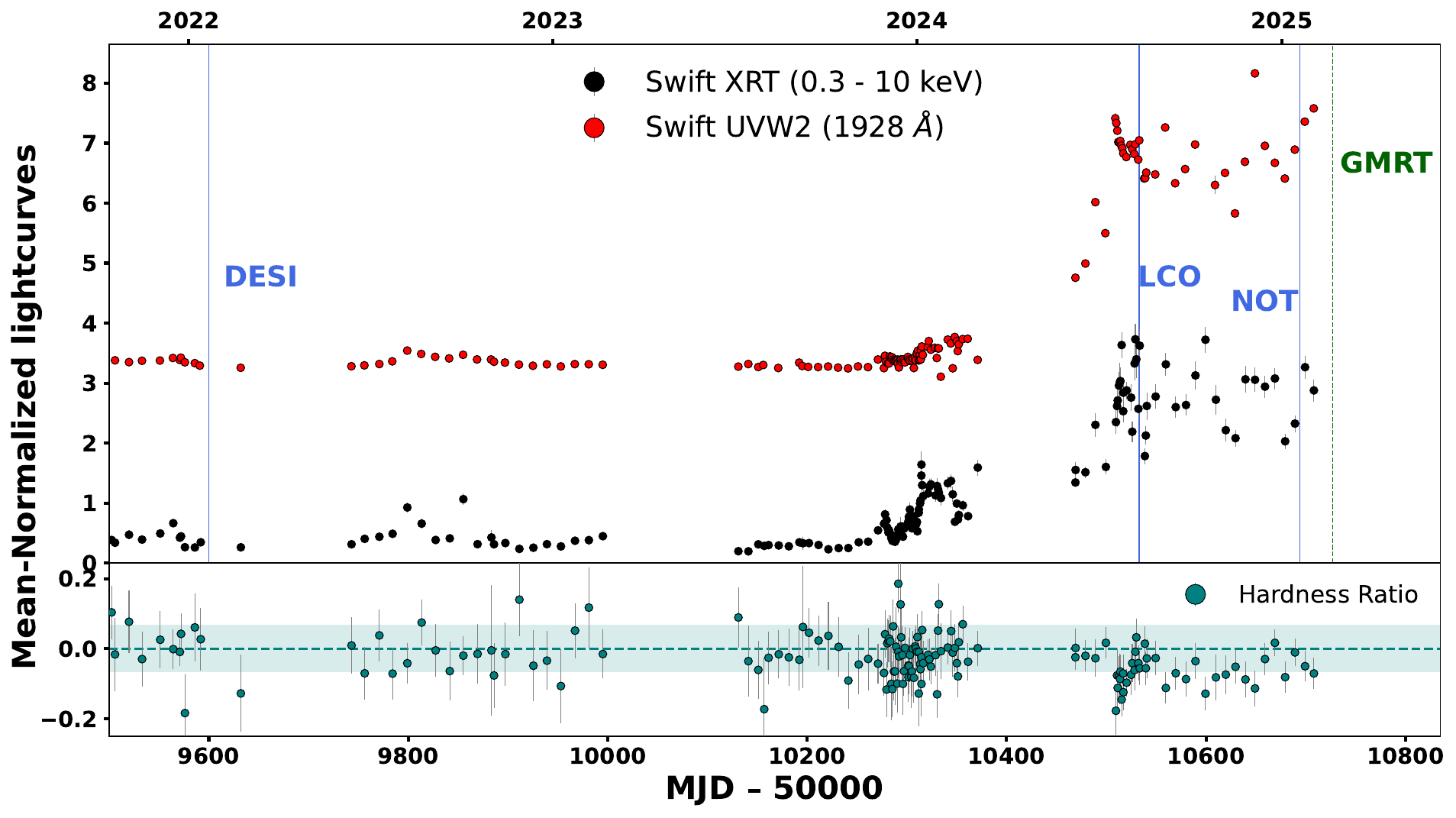}
    \caption{\textit{Top}: Multi-band light curves of Mkn~590 from 2022 to 2025, showing the pre-ignition state followed by the eventual \lq turn-on\rq\, state since August 2024. The mean-normalized light curves from \textit{Swift}-XRT and UVW2 are shown in black and red respectively, where the UVW2 lightcurve is offset by an arbitrary amount for clarity. The observational timestamps of optical spectra obtained with DESI, LCO and NOT presented in this paper are denoted by blue, solid lines. The central timestamp for  1.4~GHz GMRT observation taken in 2025 is denoted by a green, dashed line. \textit{Bottom}: A light curve of X-ray hardness ratio calculated according to the definition in Sec.~\ref{subsec:swift} along with its mean uncertainty level indicated by shaded region}.
    \label{fig:lightcrv}
\end{figure*}

Over the past 50 years, the nearby Seyfert Mkn~590 ($z$= 0.0264) has undergone multiple CL events driven by accretion rate fluctuations. In the 1970s, its spectrum was sub-type 1.5 (modest broad H$\beta$ strength of $\sim$ 2-3 $\times$ 10$^{-14}$ cgs). By the 1990s, optical continuum luminosity had increased by a factor $\sim 10$, and the spectrum was sub-type 1.0 (very strong broad H$\beta$ of flux $\gtrsim$ 10$^{-13}$ cgs). By 2014, luminosity across all bands dropped by factors of 10-100, and broad H$\beta$ vanished (sub-type 1.9; \citealt{Denney2014ApJ...796..134D}). \citet{mathur2018ApJ...866..123M}, noted the presence of Mg [II] $\lambda\lambda$2796,2803 during the low state, so at least some component of the BLR remained apparently intact and illuminated. Starting in 2017, 
Mkn 590’s luminosity at all bands has been increasing again, starting with several years of temporary multi-band flaring \citep{lawther2023MNRAS.519.3903L},
and with new broad line components emerging in the optical spectrum (VLT-MUSE, \citet{Raimundo2017MNRAS.464.4227R}; Subaru-HDS, \citet{mandal2021MNRAS.508.5296M}). 
Interestingly, the soft X-ray excess produced by the warm Comptonizing corona, 
had completely vanished between 2006--2011. However, \citet{ghosh2022ApJ...937...31G} used 
XMM-Newton to track its reappearance by 2021. The source has also been the subject of over three decades of radio monitoring. Its 1.4 GHz (centimeter or cm band) flux density declined by approximately 46\% between 1995 and the 2010s \citep{koay2016MNRAS.460..304K}. However, observations from the Very Large Array Sky Survey (VLASS) in 2021 \citep{sufia2025arXiv250701355B}  indicated that the cm-band emission has returned to levels comparable to those recorded in 1995. This recovery suggests that the cm-band variability broadly tracked the long term X-ray flux evolution, consistent with the radio/X-ray correlations observed in the low/hard states of X-ray binaries (\citealt{merloni2003MNRAS.345.1057M}). Since late 2024, Mkn~590 has been undergoing a re-brightening episode or \lq turn-on\rq\, event in the optical, UV, and X-ray wavebands, with  optical spectra transitioning from Sy~1.5 back to Sy~1.2,
marking a second CL transition \citep{2025MNRAS.540L..14P}.

In this paper, we report three new optical spectra obtained with the Nordic Optical Telescope (NOT), since the onset of Mkn~590 flux rise, and three new low frequency Giant Metrewave Radio Telescope (GMRT) observations, taken in the same observational epoch. These multi-epoch observations allow us to trace the evolution of the BLR through changes in optical emission line profiles and to characterize the current state of the cm-band radio emission.

\section{Observations}\label{sec2}

\subsection{Swift-XRT and UVOT}
\label{subsec:swift}
The \textit{Swift} telescope has been monitoring the source since 2013 and, between 2017 and 2025, observed it with a cadence of approximately 3 -- 4 days, using both the X-Ray Telescope (XRT; \citealt{Burrows2005SSRv..120..165B}) and the UV/Optical Telescope (UVOT; \citealt{roming2005SSRv..120...95R}) simultaneously. For the UVOT observations, all six filters-- V, B, U, W1, and W2 were used throughout the monitoring period. In this paper, we utilized the observations between Oct.\ 2021 and Feb.\ 2025 to construct the light curves in the X-ray and far UV band-- UVW2 ($\lambda$1928). The XRT data products were retrieved from the UK Swift Science Data Center \citealt{Evans2009MNRAS.397.1177E}. Each individual XRT spectrum was produced using a 30 arcsec circular extraction region centered on the source, with the background estimated from a co-spatial annular region spanning 35 -- 75 arcsec. The UVW2 photometric data have been reduced following standard procedure outlined by UK Swift Science Data Center where we considered a circular source extraction region of 5 arcsec and co-spatial annular background of 35 -- 75 arcsec. To account for the Milky Way extinction and host galaxy contamination in the UVW2 band, we first de-reddened the fluxes assuming the extinction law $R$=3.1 and color excess E(B$-$V)=0.0306 \citep{cardelli1989ApJ...345..245C}, and the appropriate filter functions for the UVW2 bandpass. The host galaxy contribution in UVW2 band was estimated from spectral modeling of the entire broadband spectra from the 2002 XMM-Newton observation (ObsID: 0109130301) during which the source was simultaneously observed with all six optical/UV filters and the X-ray detector. Using the \texttt{AGNSED} model \citep{kubota2018MNRAS.480.1247K} we derived a constant host-galaxy contribution of 1.55 $\times$ 10$^{-16}$ ergs s$^{-1}$ cm$^{-2}$ A$^{-1}$, which was subtracted from each UVW2 flux density measurement. A more comprehensive host-galaxy decomposition will be presented in future work. The final, mean-normalized light curves are presented in the top panel of Fig.~\ref{fig:lightcrv}. Using the \textit{Swift-}XRT count rates in the 2--10 (H) and 0.3--2 keV (S) bands, we estimated the hardness ratios (HRs) as (H$-$S)/(H+S). The mean-normalized HR light curve is presented in the bottom panel of Fig.~\ref{fig:lightcrv}.  

\subsection{Suite of optical spectra across 3 years}
\label{subsec:opt}
\begin{figure}
    \includegraphics[scale=.50]{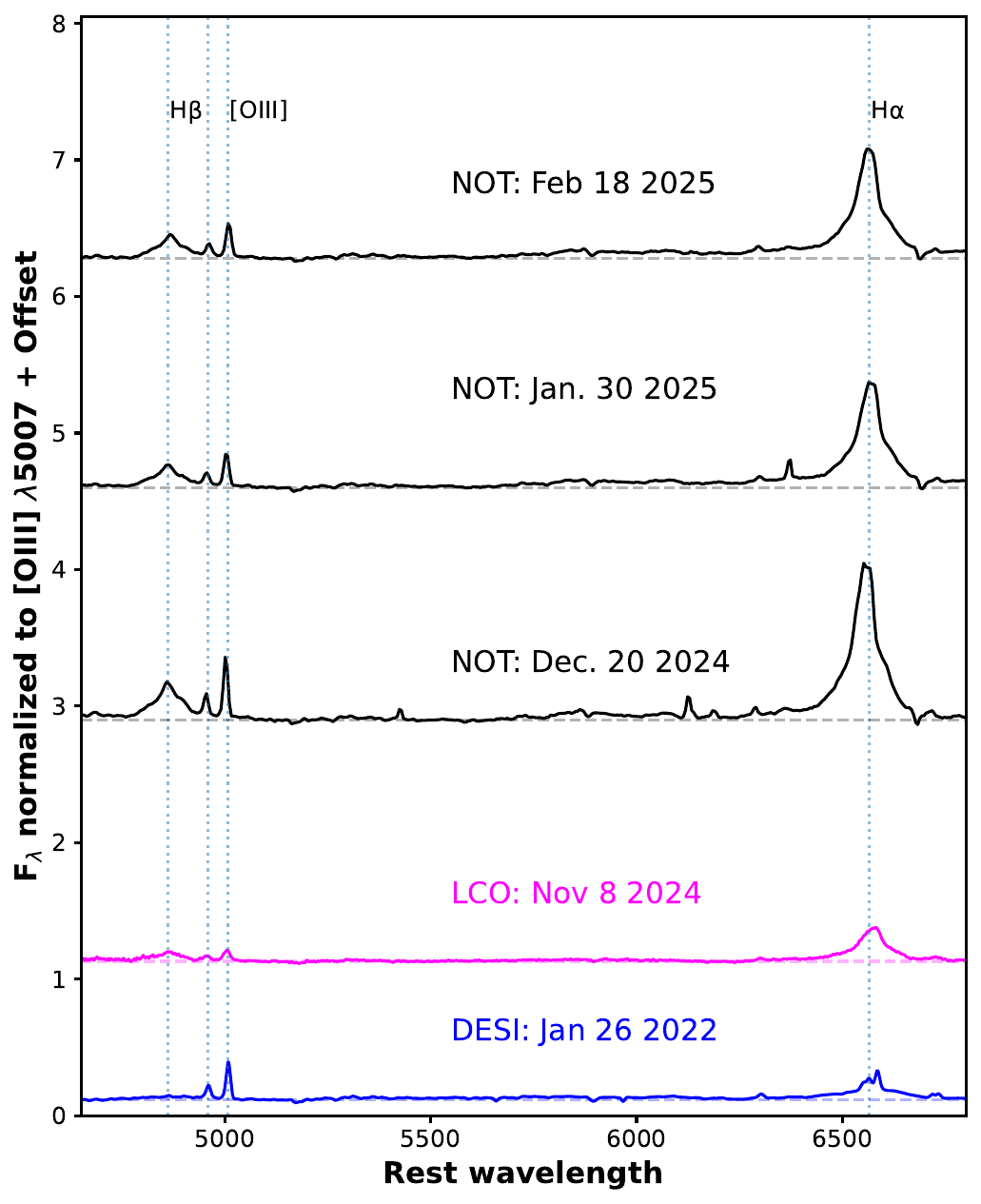}
    \caption{Optical archival spectra taken using DESI, LCO/FLOYDS, and NOT. The spectra are normalized and offset by arbitrary values for clarity. Emission lines are marked with dotted, blue lines. Dashed gray horizontal lines indicate the continuum flux level at 5100~\AA.}
    \label{fig:opt}
\end{figure}

To characterize the current state of BLR activity, we present three new optical spectra taken with the Alhambra Faint Object Spectrograph and Camera (ALFOSC) instrument of the 2.5 meter Nordic Optical Telescope (NOT; Obs ID: 70-406; PI: B.\ Palit) on 20 Dec.\ 2024, 30 Jan.\ 2025, and 18 Feb.\ 2025.  Each epoch was observed using the grism \#3 configuration with a 1'' slit, providing wavelength coverage of 3200 -- 7070 Å. The total on-source integration time across the three epochs was 3 hours. The brightness of the source, B = 14.87 and R = 14.70 mag, allowed us to obtain spectra with high signal-to-noise even under grey sky conditions. Previously, Mkn 590 was also observed by the Dark Energy Spectroscopic Instrument
 (DESI) on 26 Jan.\ 2022\footnote{the publicly available DESI DR1 spectrum was downloaded using the SPectra Analysis and Retrievable Catalog Lab (SPARCL) pipeline \citep{sparcl_2025ASPC..541...77J}.} and the FLOYDS spectrograph of the Faulkes Telescope South, through the Las Cumbres Observatory (LCO) network on 11 Nov.\ 2024 \citep{2025MNRAS.540L..14P}. The times of optical spectra from three instruments are marked in Fig.~\ref{fig:lightcrv} by vertical blue, solid lines. Next we trace the evolution of emission line profiles in the BLR within the last three years as the source varied in X-ray/UV flux.
 All spectra were flux-normalized using the [O III] $\lambda$5007 narrow emission line following the framework of \citet{vanGr1992PASP..104..700V}, adopting the line flux measured in the 2003 SDSS spectrum \citep{2025MNRAS.540L..14P} using the fitting software \texttt{PYQSOFIT} \citep{Guo2018ascl.soft09008G}. The resulting normalized spectra are shown in Fig.~\ref{fig:opt}. Going from 2022 to 2025, all lines show a clear increase in the flux of the broad H$\beta$ $\lambda$4861 and H$\alpha$ $\lambda$6563 components (indicated by vertical dotted lines) as a response to increasing X-ray/far-UV fluxes.

\subsection{Giant Metrewave Radio Telescope}
\label{subsec:gmrt}
Anticipating enhanced radio activity following the prolonged rise in X-ray flux, we submitted a Director’s Discretionary Time (DDT) proposal (ID:ddtC413; PI: B. Palit) through GMRT, located near Narayangaon, Pune in India. It consists of 30 fully steerable parabolic dishes, each 45 m in diameter, distributed over baselines of up to 25 km. We requested a total of 2.7 hours  using band~5 (1050--1450 MHz) in early 2025, that was allocated across three epochs (19 Jan., 18 Feb., and 21 Mar.). The approximate observation timestamp  is indicated by vertical green, dashed line in Fig.~\ref{fig:lightcrv}. The data analysis was performed using \texttt{CASA} version 6.7.0 \citep{CASA}. The raw data in fits format were converted to the Measurement Set format used in \texttt{CASA}. Initial flagging was performed following the standard procedure \footnote{https://gmrt-tutorials.readthedocs.io/en/latest/continuumband4.html}. The observations suffer from significant radio frequency interference (RFI), which were manually flagged. Quasars 3C 48 and 3C 63 were used as bandpass and flux calibrators, while 0217+017 was used as a phase calibrator. The delay, bandpass, flux and gain phase calibrations were carried out following the standard data reduction procedure. We performed self calibration to improve the source models and RMS noise. Imaging was performed using \texttt{tclean} with Briggs weighting and robust value of 0. The final images have been primary beam corrected. The resulting beam size for the images are: $2.39'' \times 2.23''$, $-23.14$ deg, $4.43'' \times 2.75'', 85.94$ deg, $3.51'' \times 2.92'', -84.75$ deg, respectively.
\begin{figure*}
    \centering
    \includegraphics[scale=.43]{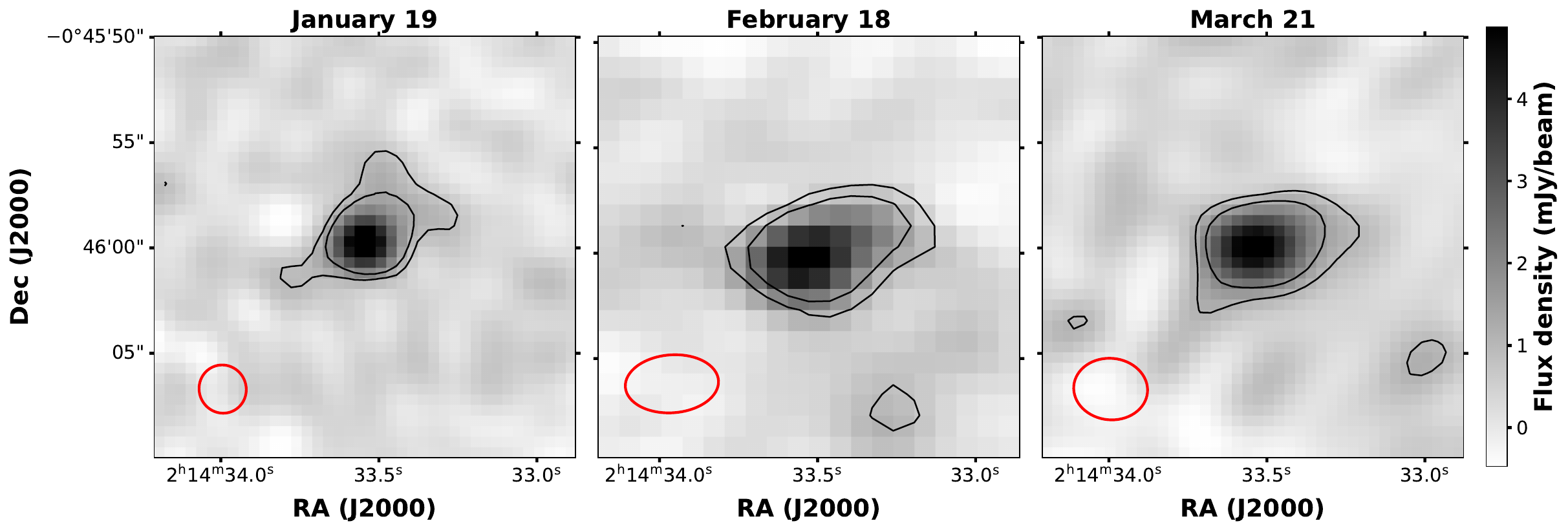}
    \caption{The three GMRT images at 1.4 GHz taken in early 2025, coinciding with the prolonged X-ray/far-UV flux rise. All images show an unresolved radio core centered at the source coordinates,  marked by black contours at 3$\sigma$ and 5$\sigma$ levels and their respective beam sizes (see Sec~\ref{subsec:gmrt}) are represented by a red ellipse in the lower left corner.}
    \label{fig:radioimages}
\end{figure*}
The images presented in Fig.~\ref{fig:radioimages} show an unresolved, core radio component, coinciding with the sky location of Mkn 590. Within a circular aperture of 5 arcsec in diameter, centered on the source, the integrated flux densities measured across three observational epochs are 7.43$\pm$ 0.30 mJy, 4.27 $\pm$ 0.30 mJy, 5.03$\pm$ 0.50 mJy. The mean flux density of 5.60 mJy. This is comparable with the 2021-2022 data from Variable and Slow Transients Survey  \citep{murphy2013PASA...30....6M,sufia2025arXiv250701355B} which report $\sim$ 5 mJy/beam at 1.3 GHz. This also agrees with archival VLA and Multi-Element Radio Linked Interferometer Network (MERLIN) observations from the mid-1980s and mid-1990s, which measured a peak intensity of $\sim$ 4.8 mJy/beam and 6.2 mJy/beam respectively \citep{koay2016MNRAS.460..304K}, when the source was in a bright-continuum, high-activity state. Compared to the 1.3 GHz VLA measurements from 2015 (\textit{L}-configuration), \citep{koay2016MNRAS.460..304K}, 
the current flux density is approximately twice as high, and significantly exceeds the European VLBI Network (EVN) high-resolution interferometric measurement also from 2015, at 1.6 GHz that detected a faint, parsec-scale radio jet component with a flux density of 1.7 mJy \citep{yang2021MNRAS.502L..61Y}.

\section{Discussion and future outlook}
\label{sec3}

In this study, we present new up-to-date optical spectral and cm-band radio observations of Mkn 590, captured during a phase of steadily increasing of X-ray/ far-UV fluxes. As shown in the bottom panel of Fig.~\ref{fig:lightcrv}, the hardness ratio estimates fall by $\sim$ 13\% below the long term average, supporting the scenario of enhanced soft X-ray emission arising from a reactivated warm corona. This interpretation is independently corroborated by count-count diagnostic plots presented in \citet{2025MNRAS.540L..14P}, where the nature of the correlated variability between hard and soft X-ray photons changed as the source brightened, indicating the emergence of additional source of soft X-ray photons. However, \citet{lawther2025MNRAS.539..501L} reported the presence of a non-negligible soft X-ray component that persisted across a range of flux states, indicating that its origin may be more complex, which requires more detailed broadband spectral modeling using physically motivated radiative transfer models of a warm corona. These will be essential to uncover the mechanisms driving this structural transformation in the inner accretion flow. This remains a goal for future work (Palit et al.\ 2026, in prep.). 

Visually, the NOT spectra obtained during this period exhibit prominently re-brightened H$\beta$ $\lambda$4861 and H$\alpha$ $\lambda$6563 components, in stark contrast to the comparatively weaker profiles observed in spectra from LCO/FLOYDS, and the near-absence of broad components in the DESI data. As shown in Fig.~\ref{fig:lightcrv}, the observed evolution closely tracks the X-ray and UV flux variations, consistent with behavior seen in earlier epochs from the 1970s to 2000s \citep{Denney2014ApJ...796..134D}. These data demonstrate that the X-rays, UV, and ionizing (EUV) continuum vary in concert, establishing that the CL-event is driven by accretion rate change, not obscuration. The NOT spectra clearly confirm that Mkn~590 is changing its sub-type. Future studies involving detailed modeling of the emission line profiles as a response to the changes in the primary broad-band continuum  will be essential to disentangle the complex co-evolution of the BLR and the accretion flow. 

Radio emission in bright Seyfert galaxies can originate from different mechanism such as collimated jets, corona, out-flowing shocks and star burst activity \citep{panessa2019NatAs...3..387P}. At few GHz, the size of variable radio emission is typically on the kilo-parsec scale. The GMRT centimeter-band observations presented in this work provide a timely snapshot of Mkn~590’s radio activity during a phase of renewed accretion. Although no significant morphological changes are observed (Fig.~\ref{fig:radioimages}), the measured 1.4,GHz flux density appears to track the long-term X-ray variability of the source. As reported in Sec.~\ref{sec3}, we detect a slight increase by a factor of $\sim2$ in the radio core emission since 2015, coincident with the source transitioning from a low to a high X-ray flux state. Following the discovery of a low-power jet \citep{yang2021MNRAS.502L..61Y}, if the optically thin synchrotron emission indeed originates from such a jet, then it is likely strongly coupled to the X-ray corona. This interpretation is particularly intriguing in light of recent work by \citet{kang2025MNRAS.tmp.1521K}, which also proposes a coupling between the X-ray-emitting hot accretion flow and weak radio jets. Together, these results reinforce the expected consistency with the fundamental plane of black hole activity -- a relationship previously demonstrated for this source during its faint state in 2015 \citep{koay2016MNRAS.460..304K}. Conversely, variable radio emission could also arise from traveling shocks or blobs, features commonly observed during transitions from the low/hard to high/soft states in X-ray binaries \citep{fender2004MNRAS.355.1105F}. As reported by \citealt{koay2016MNRAS.460..304K}, Mkn~590 exhibited a radio-to-X-ray luminosity ratio of $\sim -5$ in 2015, consistent with radio emission originating from coronal activity. Magnetically driven winds or outflows launched close to the SMBH can propagate to larger distances, potentially producing variability at parsec scales. Nevertheless, both coronal winds and a transient jets may coexist in Mkn~590.  Similar to the delayed radio response observed in the CLAGN 1ES 1927+654, we might anticipate a further rise in the cm-band emission following the recent increase in the accretion flow. Future monitoring programs will be devoted to tracking this evolution. Furthermore, distinguishing between these different scenarios will require both broadband modeling of the radio spectrum and high–angular resolution imaging, such as that provided by very long baseline interferometry, and remains one of the key goals of our future work.

We have presented new optical and radio observations of the changing-look AGN Mkn~590, captured during its ongoing re-brightening phase. The re-emergence of broad Balmer line components in the optical spectra, the softening of X-ray hardness ratio, and increased 1.4~GHz radio flux all strongly support a scenario of renewed accretion activity, encompassing a major increase in both ionizing and radio-band continuum emission as well as possible structural changes at least associated with the warm corona. Speculatively, 
transitions in accretion states may be driving structural reconfigurations in the disk-corona-jet system; this will be explored in future works. These findings reinforce the picture of coordinated variability across the X-ray, UV, optical, and radio bands, highlighting the importance of simultaneous, multi-wavelength monitoring.


\section*{Acknowledgments}
BP has been fully and AR has been partially supported by Polish National Science Center (NCN) grant 2021/41/B/ST9/04110. AB acknowledges the support of the EU-ARC.CZ Large Research Infrastructure grant project LM2023059 of the Ministry of Education, Youth and Sports of the Czech Republic. AM is supported by NCN grant 2018/31/G/ST9/03224. SP is supported by the international Gemini Observatory, a program of NSF NOIRLab, which is managed by the Association of Universities for Research in Astronomy (AURA) under a cooperative agreement with the U.S. National Science Foundation, on behalf of the Gemini partnership of Argentina, Brazil, Canada, Chile, the Republic of Korea, and the United States of America. This research uses services or data provided by the SPectra Analysis and Retrievable Catalog Lab (SPARCL) and the Astro Data Lab, which are both part of the Community Science and Data Center (CSDC) program at NSF National Optical-Infrared Astronomy Research Laboratory. NOIRLab is operated by the Association of Universities for Research in Astronomy (AURA), Inc. under a cooperative agreement with the National Science Foundation.



\bibliography{refs}%



\end{document}